\documentclass[a4paper]{article}

\usepackage[english]{babel}
\usepackage[utf8x]{inputenc}
\usepackage{amsmath}
\usepackage{graphicx}
\usepackage[colorinlistoftodos]{todonotes}

\title{The Science and Beauty of Fluidization}
\author{Entry 102371 \\ \\Frank Shaffer and Balaji Gopalan \\ USDOE National Energy Technology Laboratory}

\begin{document}
\date{}
\maketitle

Particle flows of high particle concentration are important in many fields, including chemical processing, pharmaceutical processing, energy conversion and powder transport. However, despite decades of research and industrial application, the real time behavior of particle flow fields is still not well understood. One of the main reasons is that experimental data is difficult to acquire in such harsh, opaque particle flow environments. In this educational video for the Gallery of Fluid Motion, high speed video acquired with a new, patented high speed particle imaging velocimetry (high speed PIV) technology is presented. This technology was developed by the USDOE National Energy Technology Laboratory (NETL). It is being applied to observe and measure the real time behavior of individual particle motion inside particle flow fields of high particle concentration for the first time. The high speed PIV system records high speed videos of particle motion with excellent spatial and temporal clarity. The high speed videos are analyzed to measure the concentration and the two-dimensional motion (velocity and trajectory) of individual particles. Data sample rates for velocity vectors are in the range of 0.1 to 3 million vectors per second, thereby providing full resolution of the temporal domain of particle velocity. To see and measure particle motion inside the flow fields at high particle concentrations, a custom borescope is inserted into particle flow fields. 
High speed videos and high speed PIV data have enabled careful study of the real time behavior of gas-particle flow fields. The data and insight from this new technology is proving invaluable for the design and operation of industrial systems using particle flow fields of high particle concentration. It is also proving valuable for the development of computational fluid dynamics (CFD) models of particle flow fields.

	The particle tracking technique can measure gas and fluid flow if the particles are small enough (have a low enough Stokes number) to follow the gas/fluid flow.  In areas of rapid mixing, such as turbulent wakes, particle tracking can provide better spatial resolution that conventional cross-correlation based PIV analysis.
    
keywords: fluid dynamics video, particle tracking, fluidization
\end{document}